%% file: KonradTywoniuk_QM2018_arxiv.tex
\documentclass[3p,times,procedia]{elsarticle}
\usepackage{nupha_ecrc}

\volume{00}

\firstpage{1}

\journalname{Nuclear Physics A}

\runauth{K. Tywoniuk}

\jid{nupha}

\jnltitlelogo{Nuclear Physics A}

\usepackage{amssymb}

\usepackage[figuresright]{rotating}

\input{definitions.tex}

\begin{document}

\begin{frontmatter}



\dochead{XXVIIth International Conference on Ultrarelativistic Nucleus-Nucleus Collisions\\ (Quark Matter 2018)}

\title{Quantifying jet modifications with substructure}


\author[lab1]{Konrad Tywoniuk}
\author[lab2]{Yacine Mehtar-Tani}

\address[lab1]{Theoretical Physics Department, CERN, 1211 Geneva 23, Switzerland}
\address[lab2]{Institute for Nuclear Theory, University of Washington, Box 351550, Seattle, WA 98195-1550, USA}

\begin{abstract}
The striking suppression and modification patterns that are observed in jet observables measured in heavy-ion collisions with respect to the proton-proton baseline have the potential to constrain the spatio-temporal branching process of energetic partons in a dense QCD medium. The mechanism of jet energy loss is intricately associated with medium resolution of jet substructure fluctuations. This naturally affects the behavior of the suppression of jets at high-$\pT$, inducing an explicit dependence on jet scales. In this contribution, we review recent work on using the insight from multi-parton quenching to calculate leading-logarithmic corrections to the single-inclusive jet spectrum, and discuss its impact on a wide range of observables, including jet substructure.
\end{abstract}


\end{frontmatter}














\section{Introduction}

The strong suppression of high-$\pT$ particles and jets, including heavy quarks, in heavy-ion collisions stand out as one of the hallmark measures of final-state interactions in a color deconfined medium. Strikingly, the suppression of jets persists up to very large transverse momenta. This calls into question the validity of the conventional jet quenching paradigm based on single-parton energy loss because of the large available phase space for higher-order corrections. Careful considerations of the space-time evolution of jets, see \cite{Dokshitzer:1991wu}, turns out to play an important role.

 It has already been pointed out in Monte-Carlo studies that fluctuations related to the jet fragmentation, or substructure, are extremely important for understanding experimental data, see e.g. \cite{Milhano:2015mng,Casalderrey-Solana:2016jvj}, but until recently a first-principle understanding of these corrections was lacking. In the vacuum, higher-order corrections are typically not enhanced by large logarithms because of the cancellation of real and virtual contributions for sufficiently inclusive observables. In special cases, however, they are enhanced by the phase space where the virtual terms dominate, and the procedure that allows to account for such effects is generically referred to as a Sudakov resummation.

As a classic example, let us consider higher-order corrections to the singe-inclusive jet cross section in heavy-ion collisions. In contrast to the vacuum, we argue that energy loss effects that are induced by medium interactions will give rise to a mismatch between real and virtual emissions. This occurs because a real emission happening early in the medium is sensitive to the quenching of two particles along the length of the medium. In comparison, the virtual fluctuation is only affected by quenching of the parent parton because it remains unresolved by the medium. The consequences of such a mismatch can be easily illustrated by considering an extreme scenario where the medium absorbs all jet daughter particles, leaving the leading, most energetic branching unaffected. In this case, the higher-order correction is purely virtual and simply counts number of modes that are forbidden to occur inside the medium. In terms of formation times this amounts to $\tform < L$, where $\tform \sim 1/(\omega \theta^2)$. The number of these fluctuations is then given by the phase space volume and reads $\Pi_{\tform<L}=\frac{1}{4}\abar \log^2\pT R^2 L$ in the leading-logarithmic approximation, where $\abar \equiv \alpha_s C_R/\pi$. At large $\pT$, resumming such corrections gives rise to a jet suppression factor $R_\text{jet} \sim \exp [- \Pi_{\tform < L}]$ that can be significantly smaller than unity.

Our main message is that quenching modifies directly the yield of high-$\pT$ jets and imposes further phase space restrictions for their subsequent fragmentation, and we can work out how to deal with both in a theoretically controlled manner. Here we report on a QCD calculation \cite{Mehtar-Tani:2017web} that resums a set of logarithmically enhanced higher-order corrections accounting for finite quenching effects. We will take particular care in defining the logarithmic phase space where large corrections occur, explaining the role of color decoherence in determining it.

\section{Higher-order corrections to jet quenching}

The jet suppression factor is defined as
\beq
R_\jet = \left( \frac{\dd \sigma^\med}{\dd p_T^2 \dd y} \right)\Bigg/ \left( \frac{\dd \sigma^\vac}{\dd p_T^2 \dd y}\right) \,,
\eeq
where the $\pT$-spectrum in medium is modified according to \cite{Baier:2001yt}
\beq
\frac{\dd \sigma^\med}{\dd p_T^2 \dd y} = \int_0^{\infty} \dd \epsilon\, \Pc(\epsilon) \frac{\dd \sigma^\vac(\pT + \epsilon)}{ \dd \pT'^2 \dd y} \,,
\eeq
where $\Pc(\epsilon)$ is a generic probability distribution of emitting energy $\epsilon$ out of the jet cone. In addition to its sensitivity to the jet quenching parameter $\hat q$ and the medum size $L$, it also depends on the jet $\pT$ and cone size $R$. By approximating the steeply falling spectrum by $\dd \sigma^\vac(\pT + \epsilon) \simeq \dd \sigma^\vac(\pT) \rme^{- n\epsilon/\pT}$, the nuclear modification factor is simply related to the appropriate moment of the Laplace transform of the quenching weight, $R_\jet = \Qc(\pT)$, with $\Qc(\pT) \equiv \tilde \Pc(\pT/n)$.

For ease of explanation, let us assume that all jets are initiated by quarks at high $\pT$ and expand the suppression factor $R_\text{jet}$ in terms of the strong coupling constant, 
\beq
\label{eq:NuclearModFactorExpansion}
R_\text{jet} = \Qc_q(\pT) + \Qc^{(1)}(\pT) + \Oc(\alpha_s^2 ) \,,
\eeq
where $\Qc_q(\pT) \equiv \tilde \Pc_q(\pT/n)$ is the quenching factor of a single quark. At high-$\pT$, where radiative processes dominate energy loss out of the jet cone, it is computed by resumming multiple induced gluon emissions enhanced by the medium length. Technically, this allows to neglect any interference effects, and the probability distribution is found by solving the rate equation
\beq
\label{eq:OneProngEquation}
\frac{\partial}{\partial t}\tilde \Pc_q(\nu,t) = \gamma(\nu,t) \tilde\Pc_q(\nu,t) \,,
\eeq
up to $t= L$,where $\gamma(\nu,t) = \int_0^\infty \dd \omega \, (\rme^{- \nu \omega } -1) \dd I/[\dd \omega \dd t]$ is the Laplace transform of the splitting rate (regularized by adding virtual splittings). For our present purposes, we will approximate the rate with $\dd I/[\dd\omega \dd t] = \abar \sqrt{\hat q/\omega^3}$ that accounts for multiple, soft scattering in the medium, where 
$\hat q$ is the celebrated jet transport coefficient \cite{Baier:1996sk,Zakharov:1997uu}. For this spectrum, $\omega_c \sim \hat q L^2$ acts as a cut-off energy which, if neglected, results in a time-independent rate given by,
\beq
\label{eq:QuarkQuenchingFactor}
\Qc_q(\pT) = \rme^{- 2 \abar L \sqrt{\pi n \hat q/\pT}} \,.
\eeq
Not surprisingly, this takes the characteristic form of a Sudakov suppression factor for the {\it induced gluons}, and the exponent is nothing but the multiplicity of gluons with $\omega > \pT/n$ where virtual emissions dominate over the real ones \cite{Baier:2001yt}, see also \cite{Salgado:2003gb} for further improvements. We also point out that the regime of strong quenching, i.e. $\Qc_q(\pT) \ll 1$, arises for $\pT \ll n \abar^2 \hat q L^2$.

After this short recap, we can now turn to the question of quantifying the higher-order terms.
The first $\Oc(\alpha_s)$ correction in \eqref{eq:NuclearModFactorExpansion} reads
\beq
\label{eq:AlphaSCorrection}
\Qc^{(1)} (\pT)= \int \dd z \,P_{gq}(z) \int \frac{\dd \theta}{\theta}\, \frac{\alpha_s}{\pi} \left[\Qc_{gq}(\pT) - \Qc_q(\pT) \right] \,,
\eeq
where $P_{gq}(z)$ is the Altarelli-Parisi splitting function. The first term describes the real gluon emission and its subsequent quenching, while the second term describes a virtual fluctuation where only the parent quark is affected by energy loss. In the large-$N_c$ limit, the quenching of a pair of partons is simply the combined effect of the quenching of the total charge, that is related to the color charge of the parent parton, and the additional quenching related to the additional color charge generated in the splitting \cite{Mehtar-Tani:2017ypq}. In Laplace space we can simply write $\Qc_{gq}(\pT) =\Qc_{q}(\pT) \Qc_\text{sing}(\pT) $, and the quark quenching factor that is common in both terms in \eqref{eq:AlphaSCorrection} can be factored out. While Eq.~\eqref{eq:AlphaSCorrection} describes a $\Oc(\alpha_s)$ correction, it could become sizable in a region of phase space where the quenching affects the color singlet dipole. Let us therefore proceed with a brief discussion of the singlet quenching weight $\Qc_\text{sing}(\pT) \equiv \tilde \Pc_\text{sing}(\pT/n)$.

It was shown \cite{Mehtar-Tani:2017ypq} that the resummation of multiple induced gluons for the probability of energy loss off a color singlet dipole involves both direct and interference terms. The resulting rate equation reads,
\beq
\label{eq:TwoProngEquation}
\frac{\partial}{\partial t} \tilde \Pc_\text{sing}(\nu,t) = \gamma_\text{dir}(\nu,t) \Pc_\text{sing}(\nu,t) + \gamma_\text{int}(\nu,t) \Sc_2(t) \,.
\eeq
In this case, the rate of direct emissions is identical to the one in \eqref{eq:OneProngEquation}, $\gamma_\text{dir}(\nu,t) = \gamma(\nu,t)$. The interference term is simply $\gamma_\text{int}(\nu,t) = - \gamma(\nu,t)$ for soft gluons, due to color charge conservation. However, the interference term involves a dipole suppression factor describing the survival probability of color coherence at a given time in course of the dipole propagation. It is called the decoherence parameter \cite{MehtarTani:2010ma,MehtarTani:2011tz}, and reads
\beq
\label{eq:decoherence-parameter}
\Sc_2(t) 
= \exp \left(-\frac{1}{12} \hat q \theta^2 t^3 \right) \,,
\eeq
which gives rise to a characteristic time-scale for decoherence \cite{CasalderreySolana:2011rz,MehtarTani:2011gf,MehtarTani:2012cy}.
This time-scale can easily be estimated by comparing the medium resolution scale due to multiple scattering $\lambda_\perp \sim (\hat q t)^{-\onehalf}$ with the size of the dipole $x_\perp \sim \theta t$ where $\theta$ is the dipole angle. The two scales become comparable at $\tdecoh \sim (\hat q \theta^2)^{\onethird}$, which is the so-called decoherence time.

For small angle dipoles, color decoherence takes a long time. In particular, for $\tdecoh \gg L$ or $\theta \gg \theta_c \sim (\hat q L^3)^{-1/2}$, the singlet dipole does not lose energy $\Qc_\text{sing}(\pT)|_{\tdecoh\gg L} \approx 1$. This is a manifestation of  color transparency. Hence, logarithmic corrections in \eqref{eq:AlphaSCorrection} will only arise as long as $\tdecoh \ll L$, where the singlet quenching factor becomes the product of the independent quenching factors off the dipole constituents, $\Qc_\text{sing}(\pT) |_{\tdecoh\ll L} \approx \Qc_q^2(\pT)$. Furthermore, we have to demand that the jet is formed sufficiently early in the medium so as not to be interfering with induced emissions, in particular $\tform < \tdecoh$. It can be shown that these considerations capture the leading-logarithmic behavior of the cross section and that a more sophisticated treatment of the phase space leads to sub-leading logarithmic corrections \cite{Mehtar-Tani:2017web}.

Returning to Eq.~\eqref{eq:AlphaSCorrection}, it can now be simplified as
\begin{align}
\label{eq:AlphaSCorrection_2}
\Qc^{(1)} (\pT) &\simeq \Qc_q(\pT) \times \abar \int\limits_{\tform < \tdecoh <L} \frac{\dd z}{z} \frac{\dd \theta}{\theta}\, \left[\Qc^2_q(\pT) - 1 \right] \,, \\
& \label{eq:AlphaSCorrection_3}\simeq \Qc_q(\pT) \times \left[ - \abar \log \frac{R}{\theta_c} \left(\log \frac{\pT}{\omega_c} + \frac{2}{3} \log \frac{R}{\theta_c} \right) \right] \,,
\end{align}
for fixed coupling. In going to the second line, we have focussed on the strong quenching regime, i.e. where $\Qc(\pT) \ll 1$, where only the virtual term survives. As becomes clear from \eqref{eq:AlphaSCorrection_3}, the correction is enhanced by large logarithms of the phase space related to the jet scales $\pT$ and $R$. In particular, the enhancement is single-logarithmic in jet $\pT$ because of the finite resolution angle $\theta_c > 0$, and in contrast to the toy-model considered in the Introduction due to the difference of relevant phase space.

Remarkably, in the large-$N_c$ approximation, all higher order terms in the jet suppression factor are directly proportional to the quenching of the total color charge, in the same way as in \eqref{eq:AlphaSCorrection_3}. We can therefore show that
\beq
\label{eq:jet-suppression-full}
R_\jet = \Qc_q(\pT) \times \Cc(\pT,R) \,,
\eeq
which is the main result of our analysis.
Here, $\Cc(\pT,R)$ is a novel ``collimator'' function \cite{Mehtar-Tani:2017web} that accounts for the quenching of higher-order jet fluctuations due to the mismatch of real and virtual contributions. In the strong quenching regime, the resummation of all-orders simply amounts to the exponentiation of the first-order correction in \eqref{eq:AlphaSCorrection_3}, leading to
\beq
\Cc(\pT,R) \simeq \exp \left[ - \abar \log \frac{R}{\theta_c} \left(\log \frac{\pT}{\omega_c} + \frac{2}{3} \log \frac{R}{\theta_c} \right) \right] \,.
\eeq
Note that while the quenching of the ``total charge'', or the initial parent quark, in Eq.~\eqref{eq:jet-suppression-full} does not {\it a priori} depend on the jet scales, the additional Sudakov suppression factor is sensitive to them.
We have also generalized this procedure for finite quenching effects, and devised a general resummation formula based on Eq.~\eqref{eq:AlphaSCorrection_2} for the ``collimator'' function that also goes beyond the leading-logarithmic approximation \cite{Mehtar-Tani:2017web}. 

The developments we have described in these proceeding can be employed in a phenomenological analysis of experimental data on jet suppression. More importantly, they describe a general way of extending the analysis of energy loss processes to higher orders and gaining theoretical control of their magnitude in terms of jet and medium scale analysis. They also apply directly to substructure observables, and estimates of quenching of two subjets found by grooming, based directly on the real term in Eq.~\eqref{eq:AlphaSCorrection}, were already presented at the previous edition of this conference series \cite{Casalderrey-Solana:2017mjg}. It is also worth pointing out that a similar analysis of medium scales could affect our understanding of low-momentum fragments in jets \cite{Mehtar-Tani:2014yea,Caucal:2018dla}.
Ultimately, these developments will aid in attaining a better theoretical control for Monte-Carlo implementations and will lead to a better grip on the properties of the dense QCD medium created in heavy-ion collisions.

\end{document}

%% file: definitions.tex
\usepackage{graphicx}
\usepackage{dcolumn}
\usepackage{bm}
\usepackage{epstopdf}
\usepackage{mathrsfs}
\usepackage{amsmath,amssymb,amsfonts}
\usepackage{latexsym}
\usepackage{color}
\usepackage{slashed}
\usepackage{nicefrac}

\def\sst{\scriptscriptstyle}

\long\def\comment#1{ }

\newcommand{\onehalf}{{\nicefrac{1}{2}}}

\newcommand{\onethird}{{\nicefrac{1}{3}}}

\def\sst{\scriptscriptstyle}

\def\0{{\boldsymbol 0}}

\def\x{{\boldsymbol x}}

\def\tform{{t_\text{f}}}

\def\tdecoh{t_\text{d}}

\def\pT{p_{\sst T}}

\def\abar{{\rm \bar\alpha}}

\def\Pc{\mathscr{P}}
\def\Qc{\mathscr{Q}}
\def\Cc{\mathscr{C}}
\def\Sc{\mathscr{S}}
\def\Oc{\mathcal{O}}

\def\jet{\text{jet}}
\def\med{\text{med}}
\def\vac{\text{vac}}

\newcommand{\beq}{\begin{eqnarray}}
\newcommand{\eeq}{\end{eqnarray}}
\newcommand{\be}{\begin{eqnarray*}}
\newcommand{\ee}{\end{eqnarray*}}
\newcommand{\bel}{\begin{align}}
\newcommand{\eel}{\end{align}}

\newcommand{\dd}{{\rm d}}
\newcommand{\rme}{{\rm e}}